\documentclass[11pt, onecolumn ]{article}

\usepackage[hyphens]{url} 
\usepackage{rotating, graphicx}
\usepackage{booktabs}
\usepackage[colorlinks=true,allcolors=blue]{hyperref}
\usepackage[nameinlink,noabbrev,capitalize]{cleveref}

\usepackage{lineno}
\usepackage{authblk}
\usepackage{natbib}
\bibpunct{[}{]}{,}{s}{}{;}
\begin{document}

\title{The Hiperwall Visualization Platform for Big Data Research}
\author[1]{M. Saleem \thanks{msaleem@bellarmine.edu}}
\author[2]{Hugo E. Valle \thanks{hugovalle1@weber.edu}}
\author[1]{Stephen Brown \thanks{sbrown2@bellarmine.edu}}
\author[3]{Veronica I. Winters \thanks{veronica.i.winters@vanderbilt.edu}}
\author[1]{Akhtar Mahmood \thanks{amahmood@bellarmine.edu}}
		
\affil[ 1 ]{Department of Physics, Bellarmine University}
\affil[2]{School of Computing, Weber State University}
\affil[3]{Department of Physics and Astronomy, Vanderbilt University}
		
\date{} 

\maketitle

\begin{abstract}
In the era of Big Data, with the increasing use of large-scale data-driven applications, 
the visualization of very large high-resolution images and extracting useful 
information (searching for specific targets or rare signal events) from these images 
can pose challenges to the current video-wall display technologies. At Bellarmine 
University, we have set up an Advanced Visualization and Computational Lab (AVCL) 
using a state-of-the-art next generation video-wall technology, called Hiperwall 
(Highly Interactive Parallelized Display Wall). The 16 feet wide by 4.5 feet high 
Hiperwall visualization system consists of eight display tiles that are arranged 
in a 4$\times$2 tile format and has an effective resolution of 16.5 Megapixels. Using 
Hiperwall, we can perform interactive visual data analytics of large images by 
conducting comparative views of multiple large images in Astronomy and multiple 
data events in experimental High Energy Physics (HEP).  Users can display a single 
large image across all the display tiles, or view many different images simultaneously 
on multiple display tiles. Hiperwall enables simultaneous visualization of multiple 
high resolution images and its contents on the entire display wall without loss 
of clarity.  Hiperwall's middleware also allows researchers in geographically \
diverse locations to collaborate on large scientific experiments. In this paper 
we will provide a description of a new generation of display wall setup at 
Bellarmine University that is based on the Hiperwall technology, which is a robust 
visualization system for Big Data research.

\end{abstract}

\section{Introduction}
\label{sec:intro}
Finding new information and patterns in Big Data is the current imperative~\cite{oracle1}. 
Large-scale data driven applications are on the rise, and so is the need to extract 
useful information from terabyte and petabyte scale data sets (Big Data). Additionally, the resolution 
of large images and the vast amount of data from large scientific instruments is 
also increasing which poses visualization issues of very large images that require 
a larger viewing area to capture the details~\cite{Hu2014}. So, visualizing
multi-dimensional, time-varying data sets is both a challenge to the computational 
infrastructure and to the current display technologies. 

 In this paper we have focused on a novel visualization and display technology for 
 displaying large high resolution images from large-scale data sets. A single 
 computer monitor can constrain the way in which we can analyze large volumes 
of data events.  We can view either a low-resolution image that represents the 
entire extent of the data events and thus risk missing smaller detail, or we can 
view small portions of the data events in high resolution while losing the full 
context. In order to understand the complex Big Data analysis environment for 
extracting new information, we need advance visualization and analytics tools for 
Big Data environments~\cite{Ball2005}. 

 With Hiperwall, we can see both the broad view of the data sets and the details 
concurrently which enables the shared view of complex results. With our Hiperwall 
visualization system, we can display extremely large images, conduct interactive 
analysis of large data sets, view multiple images or parameters of large data sets, 
and conduct a comparative view of many data sets (multiple data events) at once and
stream HD videos from remote collaborators (e.g.\ from CERN)~\cite{Andrews2011}\cite{Endert2012b}. 

\subsection{Visual and data analytics with Hiperwall}
\label{subsec:aboutHiper}
\begin{figure} [!ht]
\centering
\includegraphics[width=80mm]{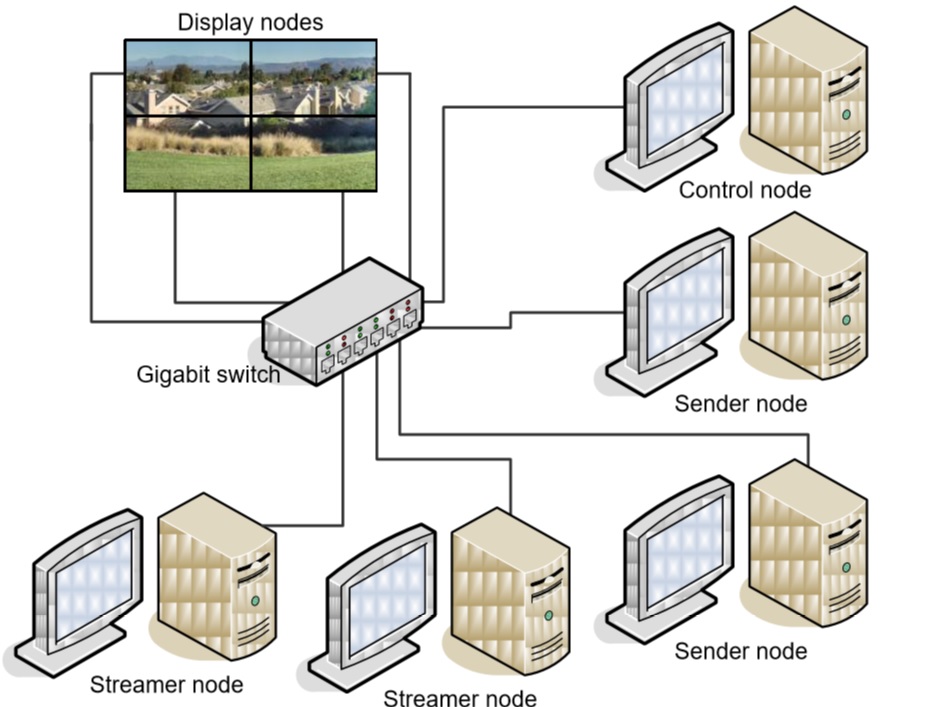}
\caption{Hiperwall terminology/topology overview\cite{hiperwallmanual}}
\label{fig:hp1}
\end{figure}

Display walls that show large quantities of information at a single glance have 
traditionally been the domain of air-traffic control centers, NASA command centers,
TV/News stations, and big budget corporations~\cite{Atlas}.  In recent years, very few such 
display walls have been installed in research labs at the large research universities.
Hiperwall started as a research project funded by a NSF grant at UC Irvine's 
California Institute for Telecommunications and Information Technology 
(Calit2)~\cite{Hiperwall}\cite{highTechStartup}.
It was designed to be a collaborative visualization platform capable of displaying 
information in real time. The 
Hiperwall software runs as an application operating on standard PCs connected 
via a standard 1 Gbps network switch. (See~\cref{fig:hp1} for  terminology/topology 
overview of the Hiperwall visualization system).

Hiperwall is a high performance, high resolution visualization system.  Hiperwall's 
visualization system software architecture applies parallel processing techniques 
to overcome the performance limitations typically associated with the display of 
large multiple images on multiple display devices (flat-panel tiles) 
simultaneously.  Hiperwall is a collaborative visualization platform that is designed to visualize enormous 
data sets and allows viewers to see the details, while retaining the context of 
the surrounding data.  This allows a group of scientists/researchers/students to collaborate and share 
detailed information. The Hiperwall system allows the user to display a wide variety 
of high-resolution 2D and 3D images, animations, movies, and time-varying data 
in real time all at once on multiple display tiles that can be configured according 
to the needs of the user.  Users can project a single image across the entire display 
area or many different images simultaneously. 
 Hiperwall's middleware even allows researchers in geographically diverse locations 
to collaborate on Big Scientific Experiments that generate Big Data.

Members of this research team from Bellarmine University have recently joined the Large Synoptic 
Survey Telescope (LSST) project and are part of LSST's Dark Energy Science Collaboration (DESC)\@.   
Equipped with a 3.2 Gigapixel camera (the world's largest digital camera), the goal of the LSST project is to 
conduct a 10-year survey of 37 billion stars and galaxies that will deliver large volumes of images and 
data sets (astronomical catalogs) that is thousands of times larger than have ever previously been compiled 
to address some of the most pressing questions about the structure and evolution of the universe, such as 
understanding the mysterious Dark Energy that is driving the acceleration of the cosmic expansion. 

\begin{figure} [!t]
\centering
\includegraphics[width=80mm]{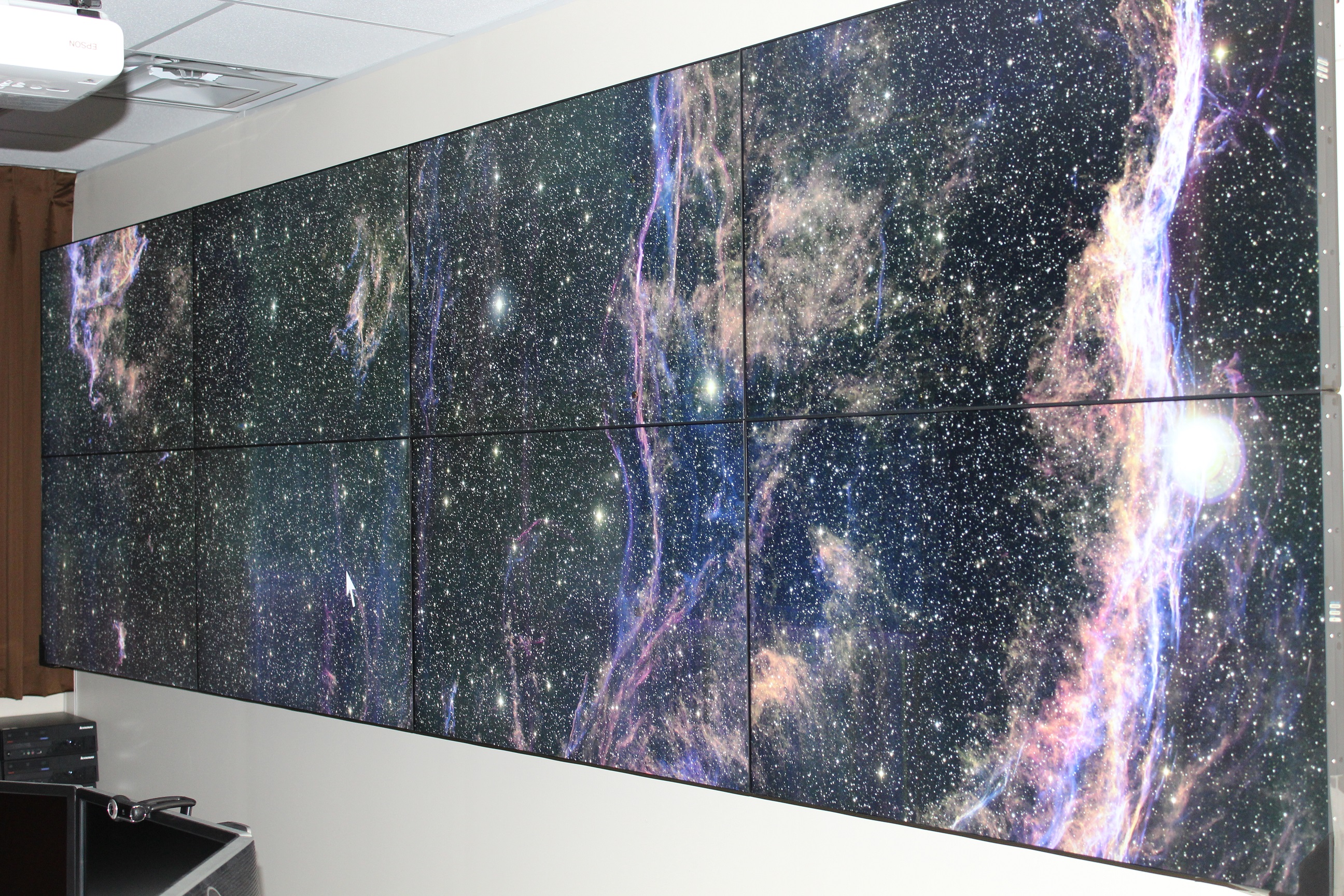}
\caption{A high resolution picture of a Supernova remnant\cite{Rector} displayed on the Hiperwall Display Wall.}
\label{fig:hp2}
\end{figure}

 LSST will produce 15 Terabytes of raw data images per night, that's about 200 
 Petabytes of imaging data over the ten years of operation, which will be the largest data set in the world~\cite{Collaboration2012}\cite{Ivezic2008}. 
Bellarmine's Hiperwall visualization system will be used for data analytics to study 
and visualize some of these large LSST images in order to extract useful and specific 
information from these large high resolution images. \cref{fig:hp2} shows a portion
of a high resolution image of a supernova remmant displayed on the Hiperwall Display 
Wall.

In the field of experimental High Energy Physics (HEP), the four big 
HEP experiments (ATLAS, CMS, LHCb, and ALICE) at the LHC (Large Hadron Collider) 
at CERN in Switzerland are producing colossal volumes of data (Big Data).  As of 
June 2017, these HEP experiments at the LHC have recorded over 200 Petabytes of 
raw data that are stored permanently at the CERN Data Center (DC)~\cite{CERN1}.
 
When proton beams collide head-on at high energies inside the LHC detectors, new 
subatomic particles are created that decay in complicated ways as they travel 
through layers of the detector.  The collision energy is converted into mass that 
create these short-lived subatomic particles. Currently, protons collide head-on 
inside the LHC detectors at an unprecedented rate of 1 billion times 
per second~\cite{CERN1}.   Most (99\%) of the raw data events are filtered out by the LHC 
experiments, and the ``interesting'' data events are stored for further analysis. The 
filtered LHC data are stored at the CERN Data Center, where the initial data 
reconstruction is performed~\cite{CERN1}.   Even after this data reduction, the CERN Data 
Center (Tier0 Grid site) processes on average about one petabyte of data per day 
the equivalent of around 210,000 DVDs~\cite{CERN3}.  In fact, in 2017, even after filtering 
out 99\% of the raw data, around 50 Petabytes of data will be stored~\cite{CERN2}.  That's 
50 thousand Terabytes of data, the equivalent to nearly 15 million HD movies~\cite{CERN2}.  
This complex data that are stored can be retrieved and accessed by thousands of 
LHC physicists worldwide for analysis via the WLCG (Worldwide LHC Computing Grid) 
and the OSG (Open Science Grid) cybeinfrastructure that is based on tiered grid 
computing technology.  On any given day, on average, more than two million grid 
jobs run on WLCG and OSG for the LHC experiments.
 
These subatomic particles that decay within the detector, leaves tracks or 
signatures of their presence that are registered by converting the particles' paths 
and energies into electrical signals to create a digital snapshot of the ``collision 
event''. High Energy Physicists have to carefully analyze these collision data 
events using dedicated algorithms, such as multivariate analysis (MVA) techniques, 
to filter out the myriad backgrounds in the collision data events and identify 
new subatomic particles (rare signal events), which could result in a new particle 
discovery.
 
\section{Hiperwall terminology and hardware components}
\label{sec:terminology}
\begin{sidewaysfigure*}
\centering
\includegraphics[width=\textwidth]{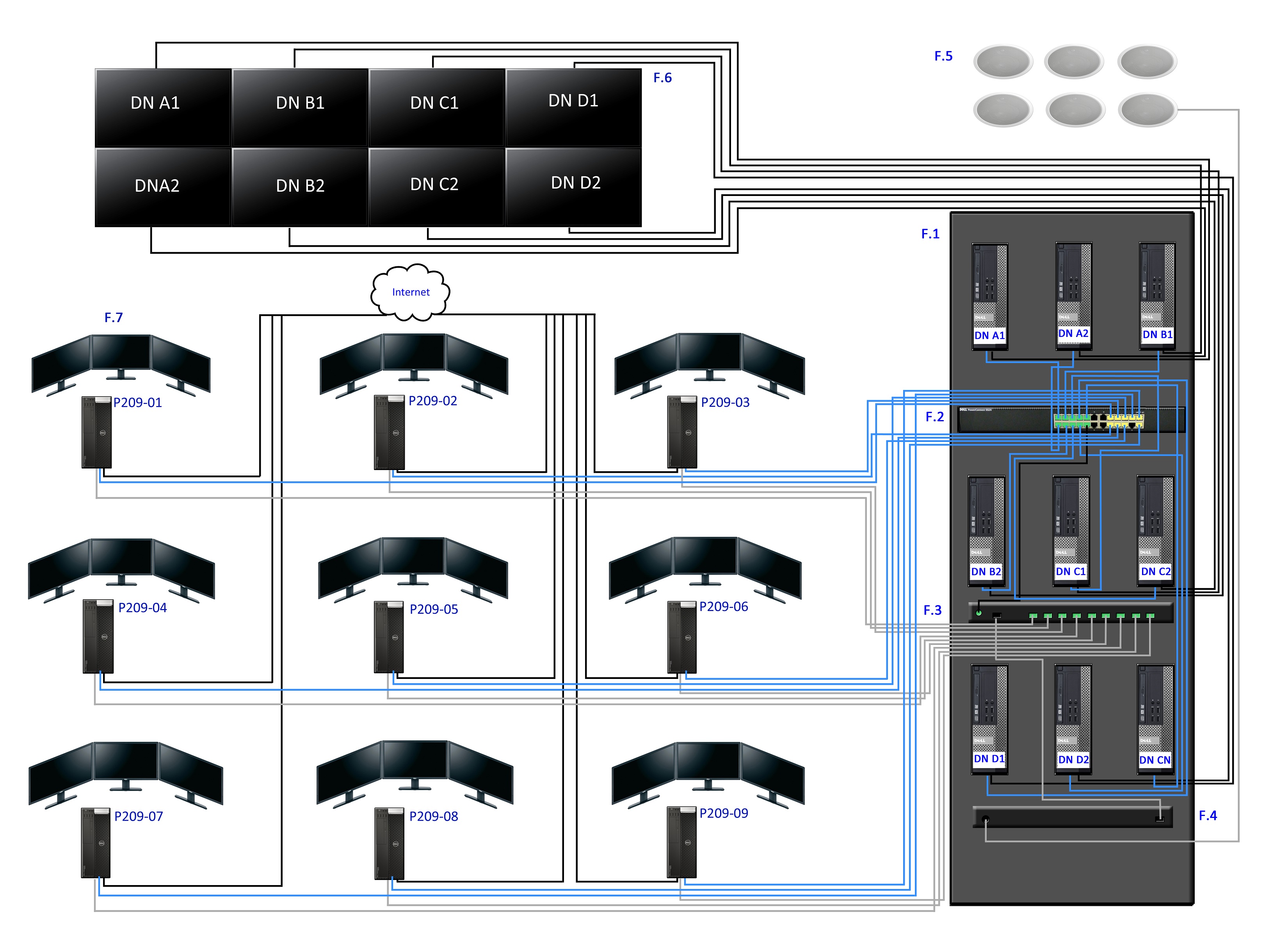}
\caption{Advanced Visualization and Computational Lab (AVCL) schematics. See~\ref{table:specs} 
for description of components.}
\label{fig:hp3}
\end{sidewaysfigure*}

In this section we describe the Hiperwall terminology/topology (as shown in~\cref{fig:hp1}), 
that includes the main Hiperwall hardware components and 
the schematics of the Hiperwall visualization system, shown in~\cref{fig:hp3}. 
Our Hiperwall display wall is connected to 9 high-end Data Analysis workstations 
equipped with dual-NVIDIA multi-core graphics technology that provides parallel 
processing capabilities needed to visualize very large data sets quickly and efficiently.  
A description of Hiperwall hardware components and specifications are listed in~\cref{table:specs}.

\begin{table*}
\centering
\begin{tabular}{lp{12cm}}
    \toprule
    \textbf{Figures} & \textbf{Hiperwall Visualization System Hardware Components and Specifications}\\ 
    \toprule
    \cref{fig:hp3}.F.1 & Hiperwall Visualization Cluster (1 Control and 8 Display Nodes)

    \textbf{A. Rack-Mounted Control and Display Nodes:} 9 Dell Optiplex 7010 PCs 
    on a 40U Rack (1 Control Node and 8 Display Nodes) 
            Specs for each node: $(3^{rd}$ Gen Intel Core 3.4 GHz i7--3770 Processor; 
            4 GB 1600 MHz DDR3 RAM\@; 1 GB RAM AMD RADEON HD 7570 Graphics Card; 500 GB 
            Hard Drive (Total RAM\@: 9 GB\@; Total Hard Drive Space of 4.5 TB).  
            Operating System: Windows 7 (64-bit).

    \textbf{B. Secondary Control Node (Not Shown in~\cref{fig:hp3}):}
	Dell Latitude 10 Tablet. Specs: 1.8 GHz Z2760 Intel Atom processor; 2 GB 
            RAM\@; 533 MHz Intel Graphics Media Accelerator Graphics Card; 2 GB DDR2 SDRAM
            RAM\@; 64 GB Hard Drive. Operating System: Windows 8 Pro (32-bit).
    \\ 
    \midrule 
     \cref{fig:hp3}.F.2 & \textbf{1 Gbps 24-port Network Switch:} 1U Rack-Mount Dell 
     1 Gbps 24-Port PowerConnect 2824 Switch.
    \\ 
    \midrule
     \cref{fig:hp3}.F.3 & \textbf{Audio Digital Signal Processor:} Biamp Nexia-CS-10
      mic/line inputs and 6 mic/line outputs.
    \\ 
    \midrule
     \cref{fig:hp3}.F.4 & \textbf{Audio Power Amplifier:} 2U Rack-Mount TOA 900 
     Series-II P-912MK2.
    \\ 
    \midrule
     \cref{fig:hp3}.F.5 & \textbf{Ceiling Speakers:} 6 Ceiling-Mounted Speakers, 
     Atlas Sound FAP62T (Strategy-II series) 6'', 32W.
    \\ 
    \midrule
     \cref{fig:hp3}.F.6 &\textbf{Display Wall:} 
        8 commercial-brand 55'' Samsung LED (backlit) HDTV (Model UD55C). Average
                Bezel size = 0.1 inch (2.75 mm).
                
       Display Wall Layout: 4$\times$2; Effective resolution (4$\times$1920) $\times$ 
                (2$\times$1080) = 7680$\times$2160 (16.5 Megapixels).
                
         Display Wall Dimensions: Width = 16 feet (4.9 m) and Height = 4.5 feet 
                (1.4 m); Bezel-to-Bezel = 0.2 inch (5.5 mm).
                
         Connection from Control/Display nodes to display wall: HDMI.
    \\ 
    \midrule
     \cref{fig:hp3}.F.7 & \textbf{Sender/Streamer Nodes (9 Data Analysis Workstations):}
 
        9 Dell T5600 Workstations each connected to three 24'' monitors.
        
        Specs for each Workstation: Each Node has a Dual-Four Core 3.3 GHz Xeon 
            E5--2643 processor; 64 GB DDR3 RDIMM RAM\@; 2 GB$\times$2 GB RAM NVIDIA Quadro 
            Graphics Cards; two CUDA-768 cores (1536 CUDA cores) per workstation; 4 TB 
            Hard Drive (Total RAM\@: 576 GB\@; Total Hard Drive Space: 36 TB).  
            Operating System: Dual-Boot Windows 7 (64 bit) and Scientific Linux 6.4.
    \\ 
    \bottomrule
\end{tabular}
\caption{Hiperwall Visualization System Hardware Components and Specifications 
    as shown in~\cref{fig:hp3}.}
\label{table:specs}
\end{table*}

\begin{figure*} [!t]
\centering
\includegraphics[width=\textwidth]{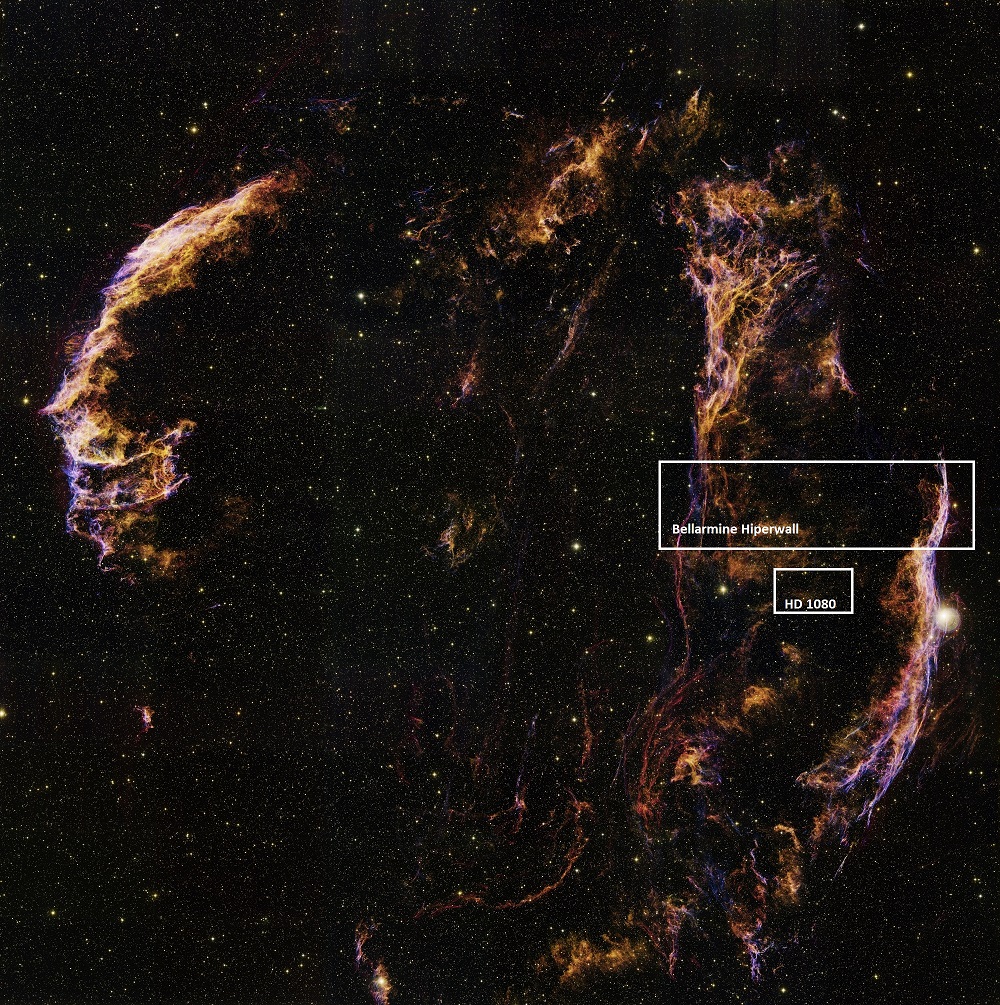} 
\caption{The total viewing area of the 7680$\times$2160 pixels (16.5 Megapixels) 
    Hiperwall Display and a standard HD 1920$\times$1080 (2 Megapixels) monitor 
    superimposed (shown as white boxes) on top of a 24457$\times$24576 pixels 
    (600 Megapixels) large astronomical image of a supernova remnant (SNR)~\cite{Rector}.
\emph{Note that a single LSST image~\cite{Hansen2011} will be about five times 
larger than this astronomical image.}}
\label{fig:hp8}
\end{figure*}

\subsection{The control node}
\label{subsec:controlNode} 
		The control node (labeled as DN CN in~\cref{fig:hp3}) is the brains behind the Hiperwall system. It gives the user the 
		ability to control what, where, how, and when the contents are displayed. The user interface
		shows a miniature view of the eight Hiperwall display tiles. The interactive Hiperwall
		drag-and-drop simplicity gives the user the ability to place contents on the display tiles.

\subsection{The secondary control node}
\label{subsec:secondaryNode}
		The secondary control node provides the user the ability to control the Hiperwall 
		system (nodes) remotely from a distant location via a portable device, 
        such as a tablet. The secondary control node is not shown in~\cref{fig:hp3}.
		
\subsection{The display node}
\label{subsec:diaplyNode} 
        The 8 display node PCs in the rack (labeled as DN A1 through DN D2) as shown in 
        \cref{fig:hp3}.F.1 are connected to the Hiperwall display tiles (one display node 
		for each display tile). Each display node consists of a display device that is driven by 
		a PC (mini-tower or rack-mount) running the Hiperwall software. The Hiperwall software runs
		as an application operating on a standard PC connected 
		via a standard 1 Gbps network switch. Hiperwall's tiled display system 
		software architecture applies parallel processing techniques to overcome the performance
		limitations typically associated with the display of large multiple images on multiple
		display devices (flat-panel tiles) at once. A display node software 
		receives content over a 1 Gbps network switch from a sender or a streamer node and
		displays it on wall tile via a HDMI cable. All display-node computers work in parallel, thus giving 
		flexibility and scalability while also allowing the user to place contents on one or 
		more multiple display tiles all at once.
\begin{figure} [!ht]
\centering
\includegraphics[width=80mm]{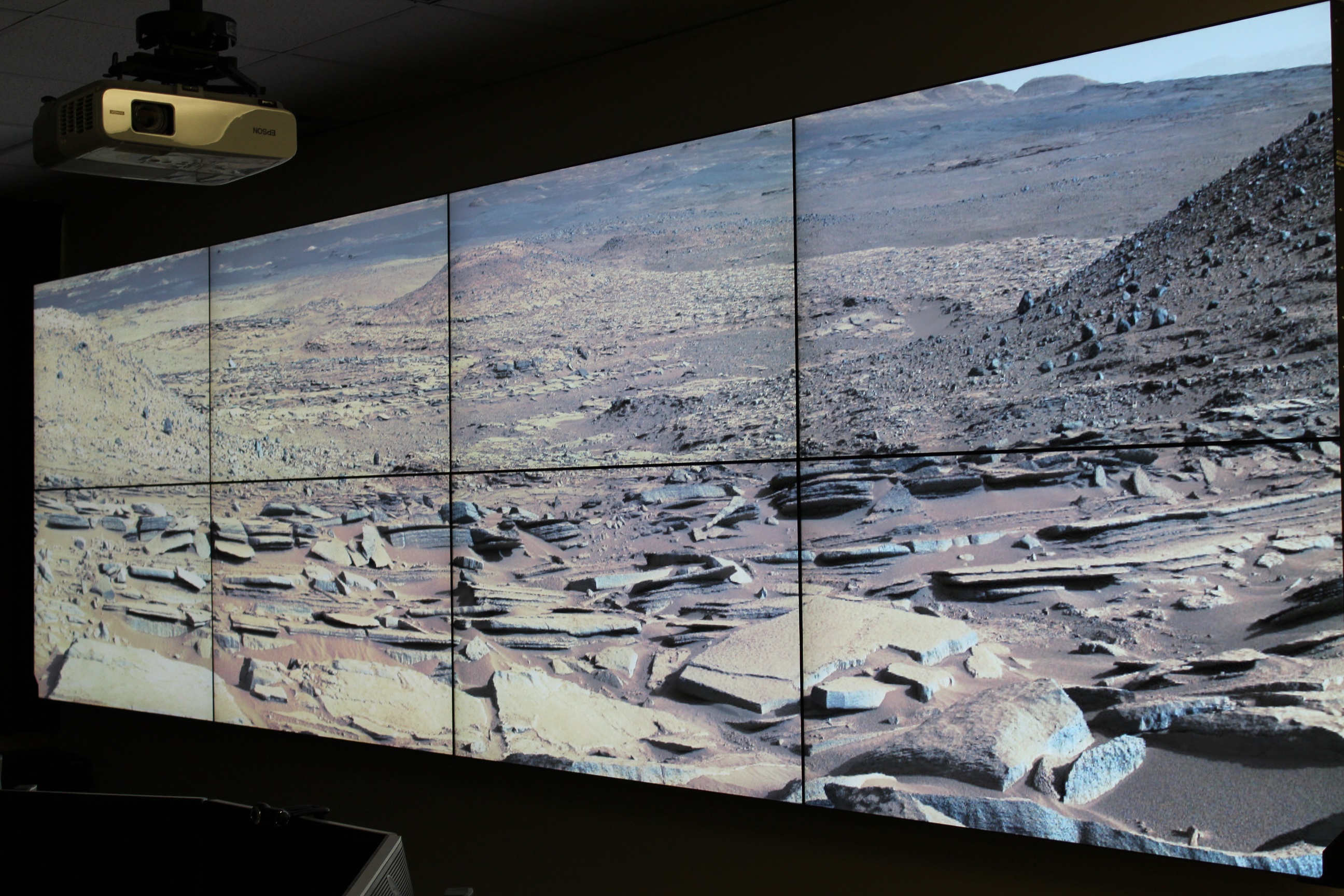}
\caption{A high resolution picture of the surface of Mars taken by NASA's 
Curiosity rover displayed on the Hiperwall Display Wall\cite{Nasa}.}
\label{fig:hp4}
\end{figure}

\subsection{The display wall}
\label{subsec:displayWall} 
    Each of the 8 display tiles (labeled as DN A1 through DN A2) as shown in 
    \cref{fig:hp3}.F.6 are connected to the corresponding display node PC 
    in the rack.  The display wall displays the contents from the user's Workstation (sender and 
    streamer nodes) onto the display wall (flat-panel tiles) via the control node. Each tile 
    is a high-end commercial-brand 55'' (UD55C) Samsung LED (backlit) HDTV with a 
    very narrow bezel size of about 2.75 mm (0.1 inch). Actually, the upper and the 
    left bezel size is 3.7 mm (about 0.15 inch), whereas the bottom and the right bezel 
    is 1.8 mm (about 0.07 inch). Each of the display node PCs is responsible for displaying and 
    and rendering a portion of the overall image that is being shown/displayed on the Hiperwall. Thus,
    all the PCs work in parallel to render the total image. 
    
     The Hiperwall software runs on each 
    display node and renders images though the control node which transfers the data displayed 
    on Hiperwall display tiles via display nodes.  The Hiperwall system provides robust 
    visualization of multiple forms of display and monitoring capabilities simultaneously 
    with a lot of user flexibility. With the Hiperwall visualization system we 
    can display various types of images (1 GB or larger) and contents on the display 
    wall.  We can add, resize, scale, rotate, zoom and reposition large images, 
    charts, plots and HD video feeds for visualization, simultaneously.  Users can 
    spread one large piece of image or spread a variety of images across the entire display wall.  In our setup, the 8 Hiperwall
    display tiles are arranged in a 4$\times$2 format. The display wall is 16 ft 
    wide by 4.5 ft high with an effective bezel-to-bezel size of only 0.2 inch (5.5 mm). 
    The effective resolution is 7680$\times$2160 pixels which is 
    about 16.5 Megapixels. \cref{fig:hp8} shows an actual 24457$\times$24576 pixels
    (600 Megapixels) large astronomical image of a supernova remnant. In comparison,
    the total viewing area of our Hiperwall Display Wall and a standard HD 1920$\times$1080 
    (2 Megapixels) monitor are shown as white boxes. A single LSST image~\cite{Rector} 
    will be about five times larger than this astronomical image. 

\subsection{The sender/streamer nodes}
\label{subsec:senderStreamerNode}
    The sender and streamer nodes deliver content to the Hiperwall display tiles 
    via the 1 Gbps network switch. The sender/streamer nodes are high-end workstations. It is called a 
    sender node if it is running the Hiperwall sender software license for displaying still
    images and animations. A streamer node runs the Hiperwall streamer software license for displaying
    HD videos. Sender nodes allow users to view the contents of the monitor(s) connected to the 
    users' workstation. The sender/streamer nodes can run multiple applications of the 
    Hiperwall sender/streamer software. A Java application allows the output of 
    the user's application to be transported across the network for viewing
    on the Hiperwall display wall tiles. Users can send the entire screen to the display wall or 
    divide it up into rectangular regions and deliver each region to the Hiperwall as an independent
    object. \cref{fig:hp4} shows a high resolution picture of the surface of 
    Mars taken by NASA's Curiosity rover that was displayed/projected 
    from one of the sender nodes (workstation) onto the Hiperwall Display Wall.
    
     The screen sender software can also run remotely. At Bellarmine, we have 11 sender licenses and 1 streamer
    license, allowing us to have up to 12 display projections at a time. Streamer software can 
    provide high frame rate feeds to the Hiperwall displays and are used for HD videos and animations. The
    streamer system can stream up to 60 frames per second (FPS) at a resolution of 1080P or higher. FPS
    is the frequency (rate) at which an imaging device produces consecutive images. 60 FPS is typically
    used because high definition televisions (HDTV) are designed around 60 Hz signals as a common
    base capability. 60 FPS is optimal to the eye and gives a much more realistic sense of motion
    for what is happening on the screen~\cite{Hdtv}. A high resolution image or HD video feed can be 
    enlarged and stretched across the entire display wall or multiple arrangements of 
    many images or HD video feeds can be arranged in any size or position across the 
    display tiles and zoomed without loss of clarity. 

\begin{figure} [!b]
\centering
\includegraphics[width=80mm]{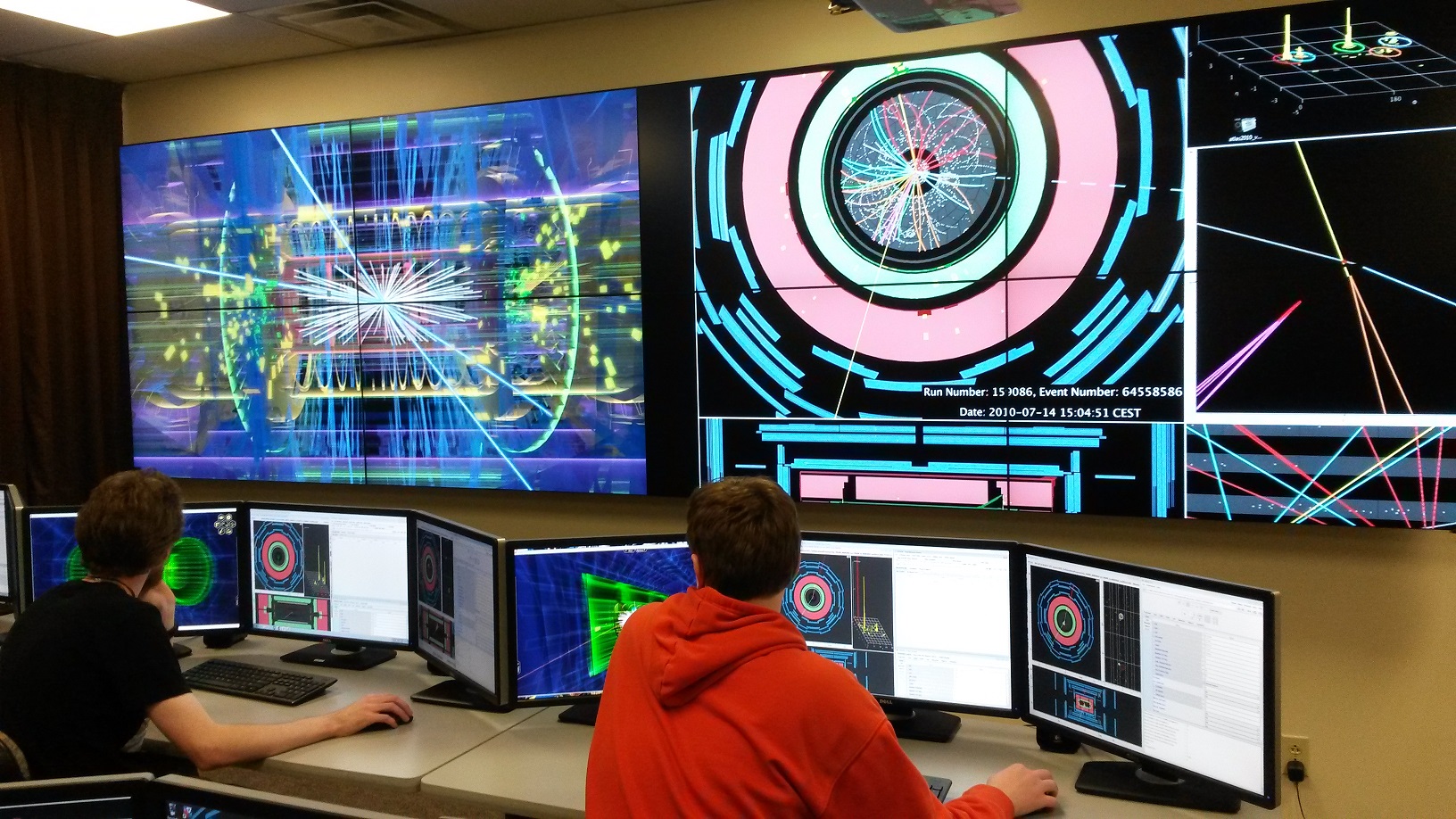}
\caption{Students analyzing particle tracks from High Energy Physics (HEP) 
proton-proton collision data events.}
\label{fig:hp5}
\end{figure}

\section{Collaborative research and learning environment} 
\label{sec:collaborativeLearning}
Hiperwall display wall system provides a dynamic and interactive learning experience for students. Hiperwall 
technology lets teachers, students, and researchers view large amounts of information using PCs via 
a regular Ethernet connection. Students can also connect from laptops or tablets to view information. Teachers
can better engage with students, turning traditional classroom lectures into an interactive, collaborative, 
and enriching experience. Hiperwall technology allows students, teachers, and researchers to visualize large
Terabyte-size data sets with high resolution and to conduct interactive visual analytics of large data sets. 

 We have written a User's Manual for the Bellarmine Hiperwall Visualization 
System with input from undergraduate students who are our primary users.  We have 
also designed and developed a web-portal interface that allow users to select 
specific display-tile arrangements for visualization at the click of a button, 
based on the what is displayed on the user's Data Analysis Workstation's 
(sender/streamer nodes) triple monitors. 

\section{Conclusion}
\label{sec:conc}
\begin{figure} [!t]
\centering
\includegraphics[width=80mm]{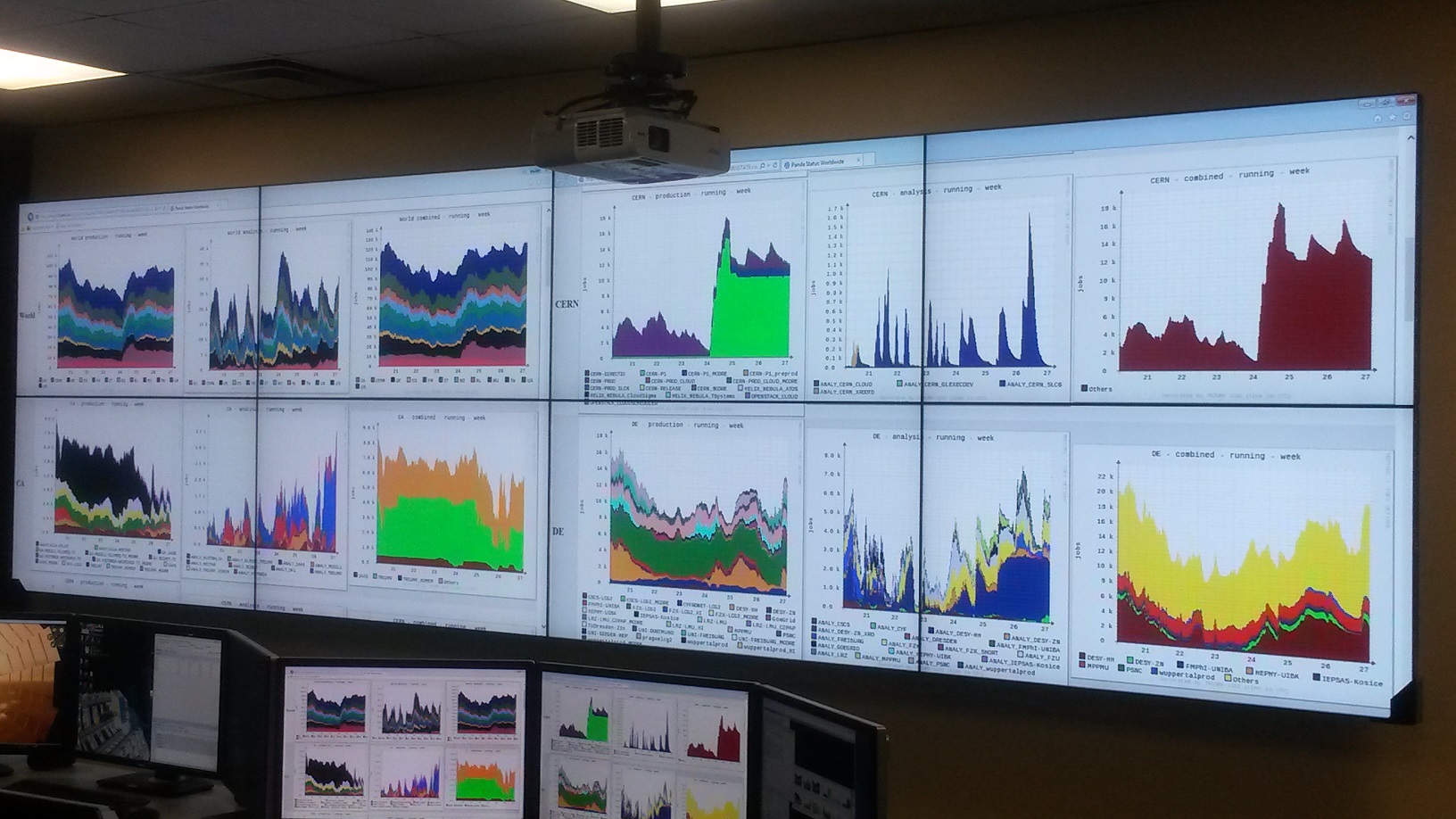}
\caption{Visual Analytics using Hiperwall Display Wall to monitor grid jobs.  }
\label{fig:hp7}
\end{figure}

We have described the Hiperwall visualization system setup at Bellarmine University to 
display large images in Astronomy to enhance the display size and for interactive 
data analysis for HEP (High Energy Physics) data events from the ATLAS experiment 
at the LHC (Large Hadron Collider) at CERN\@. \cref{fig:hp5} shows students 
in the AVCL Lab analyzing particle tracks 
from the High Energy Physics (HEP) proton-proton collision data events. As shown in 
\cref{fig:hp7}, we have also used the Hiperwall Display Wall for visual analytics 
to monitor grid jobs. We broke new ground by being the first 
undergraduate institution in the US (and the only institution in Kentucky) to implement 
the Hiperwall technology in the field of High Energy Physics.

 The Hiperwall technology has an architectural advantage since Hiperwall runs
on standard PCs connected to standard Ethernet network, thus eliminating the need for  
specialized switches and servers.  Each display node renders its own portion 
of the display, thus eliminating the bottlenecks associated with central servers 
that make Hiperwall a robust visualization system. The key breakthrough of the 
Hiperwall system is the distributed software architecture that overcomes 
performance limitations typically associated with the display of numerous 
large contents on multiple display tiles. This paper suggests a cost-effective solution for visualizing high resolution 
large-scale data-driven images and applications with a broad view using the Hiperwall 
visualization system based on the setup at Bellarmine University's Advanced 
Visualization and Computational Lab (AVCL)\footnote{It must be noted that purpose 
of the Hiperwall is not about increasing resolutions. 
Once the original image has been created the resolution can not be changed. The Hiperwall 
can increase image size by adding pixels by interpolation methods, but the original 
information is not increased. Therefore, the purpose of the Hiperwall is to be able 
to show detail while showing more area.  The user gains more visible area of the 
image without losing detail.}.

 The Hiperwall visualization system is an effective way to display a large 
image or multiple images on display tiles. We have also used the Hiperwall visualization 
system successfully for conducting workshops on virtual Physics Labs as well as 
interactive workshops on Python in a classroom setting. So we have demonstrated that
 the Hiperwall visualization system can 
also be used as a collaborative and interactive learning environment in classrooms. 
Hiperwall visualization system is also useful for collaborative research work with 
other institutes by allowing researchers, teachers, students to visualize and analyze 
the data with high resolution and  to conduct interactive analysis for large data sets over the internet.

\begin{figure} [!h]
\centering
\includegraphics[width=80mm]{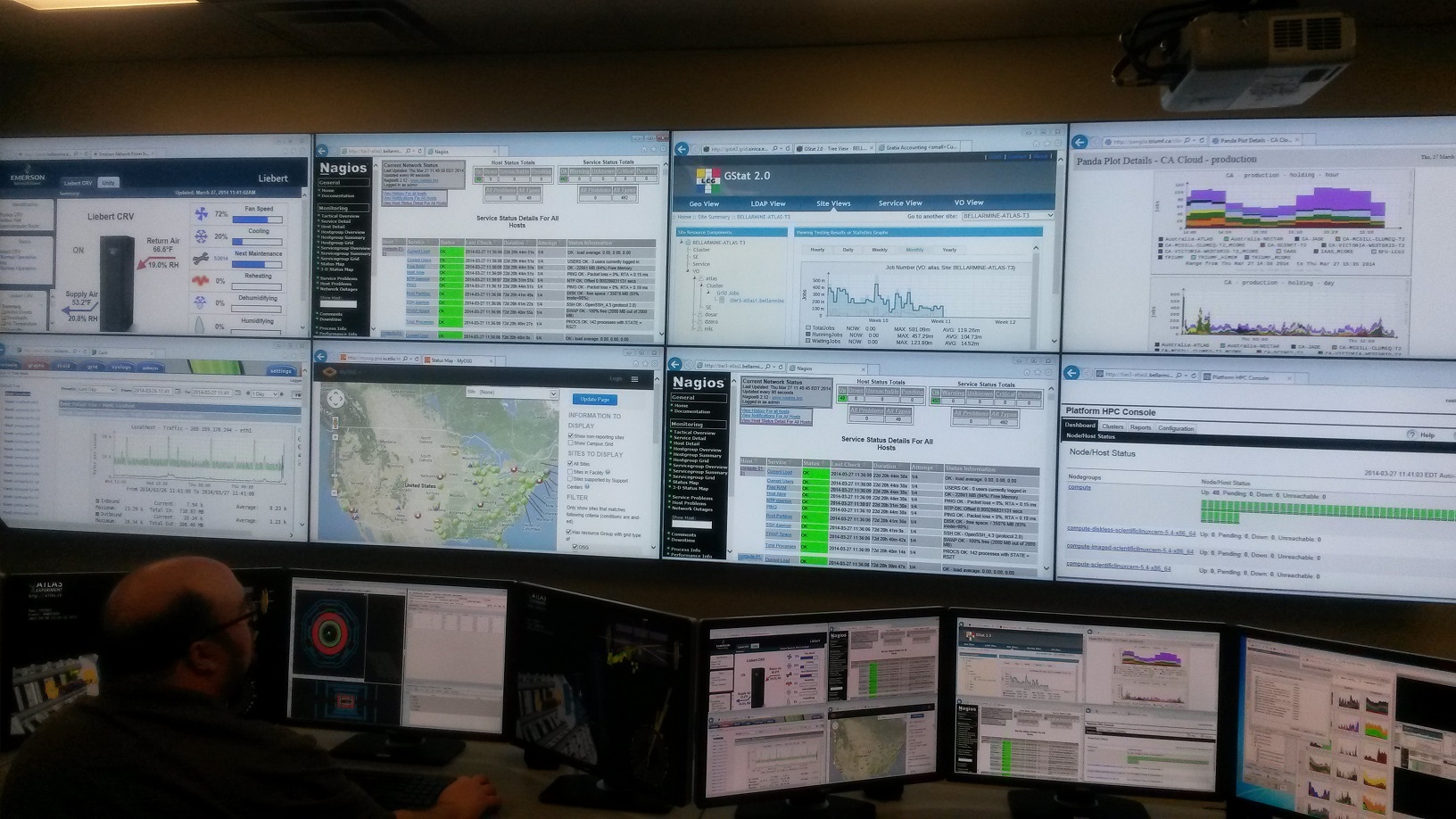}
\caption{Hiperwall Display Wall being used as a Open Science Grid (OSG) Tier2 Grid Operations Center at AVCL\@.}
\label{fig:hp6}
\end{figure}

 The Hiperwall is an effective visualization system for Big Data applications, 
telecommunications, and for conducting visualization studies of large-scale images 
and data events from Big Science experiments in experimental High Energy 
Physics (HEP), Astrophysics, and Astronomy.  Hiperwall Visualization System can 
also be used as a Operations/Command center for grid sites on OSG (Open Science Grid), 
like the NASA mission Operations/Command center. At Bellarmine University we have 
also used the Hiperwall Visualization System as a Tier2 OSG Grid Operations/Command Center, 
as shown in~\cref{fig:hp6}.

\section*{Acknowledgements}
The authors would like to thank the National Science Foundation (Grant \#1229306
and \#1154454) and the Henry Luce Foundation (Clare Boothe Luce Program) 
for supporting this research project.


\bibliography{mybib}

\end{document}